\documentclass[aps,prl,preprint,groupedaddress,nofootinbib]{revtex4-1}
\usepackage{graphicx}
\begin{document}


\title{\large  
Affleck-Dine Leptogenesis with One Loop Neutrino Mass and strong CP
}


\author{\bf Rabindra N. Mohapatra$^a$}
\author{Nobuchika Okada$^b$}
\affiliation{}
\affiliation{$^a$ Maryland Center for Fundamental Physics and Department of Physics, University of Maryland, College Park, Maryland 20742, USA}
\affiliation{$^b$ Department of Physics, University of Alabama, Tuscaloosa, Alabama 35487, USA}


\date{\today}

\begin{abstract} 
We present a unified model that solves  four major problems of the standard model i.e. neutrino masses, origin of matter, strong CP problem and dark matter using the framework of Affleck-Dine (AD) mechanism. The AD-field is responsible for inflation, origin of matter and neutrino masses which arise at the one loop level. Neutrino masses are therefore intimately connected to the baryon to photon ratio of the universe.  The dark matter in the model is the axion field used to solve the strong CP problem.  
The model has a near massless Majorana fermion which contributes to $\Delta N_{\rm eff}\sim 0.1$ in the early universe, that can be tested in the upcoming CMB-S4 experiment. 

\end{abstract}

\maketitle

\section{1. Introduction} 
The standard model (SM) despite its phenomenal experimental success is an incomplete model. 
Its major deficiencies that call for extension are  its inability to explain three experimental  observations : (i) small neutrino masses; (ii)  matter-anti-matter asymmetry in the universe; (iii) the dark matter of the universe. 
A fourth theoretical problem with the SM is why strong CP violating parameter $\theta$ is so small (i.e.~$\theta \leq 10^{-10}$). 
In an attempt to address the first three of these problems, we recently proposed an extension of the SM \cite{MO1} 
using the framework of the Affleck-Dine (AD) mechanism \cite{AD} for leptogenesis. In this model, a complex scalar field, called AD field here, generates the lepton asymmetry as it evolves from the early stage of the universe.
Our model~\cite{MO1}  provides an example of how to implement leptogenesis in a minimal model with radiative neutrino masses. The AD field, also played the role of inflaton whose non-minimal coupling to gravity leads to a viable model of inflation in the early universe. Thus the AD field played a key role in not only implementing leptogenesis but also in generating neutrino masses as well as the inflationary expansion
In this paper, we show how a similar but a more economical version of the model in Ref.~\cite{MO1} can provide an axion solution to the strong CP problem. 

We work within the invisible axion model  framework \cite{K, SVZ, DFS, Z}, of KSVZ type, where
 the Peccei-Quinn (PQ) symmetry breaking scale  is in the range of $10^9-10^{12}$ GeV as required by astrophysical considerations.  The PQ symmetry breaking also provides a lepton number breaking term involving the AD field which is crucial to AD leptogenesis.

Our starting point is how to implement leptogenesis in minimal models for small neutrino masses. As is well known, connecting the origin of neutrino masses to the matter-antimatter asymmetry via the mechanism of leptogenesis~\cite{FY} is an attractive possibility and has been the subject of great deal of activity over the past decades~\cite{rev1,rev2}. However, this connection  is most compelling only for the case of type I seesaw mechanism~\cite{seesaw1,seesaw2,seesaw3,seesaw4,seesaw5} with two or three right handed neutrinos. 
On the other hand, there are other very interesting mechanisms for generating small neutrino masses, such as type II, type III and inverse seesaw as well as loop models (see for some example of loop models \cite{zee,chang, babu,salah} and an exhaustive review in Ref.~\cite{volkas}). 
In the latter class of models,  it becomes necessary to add extra particles  to implement leptogenesis. These extra particles do not have anything to do with neutrino mass generation but are put in solely to implement leptogenesis.  For a  discussion of traditional  leptogenesis and the need for extra particles, 
see Ref.~\cite{hambye}  for  type II seesaw, Refs.~\cite{haba, agashe} for inverse seesaw, and  Ref.~\cite{Gu} for loop models.
 For one class of loop models for neutrino masses, we showed in Ref.~\cite{MO1} that use of AD mechanism provides a way to avoid adding extra particles to generate the lepton asymmetry., which  in combination with the sphalerons,  leads to baryon asymmetry of the universe~\cite{KRS}.  (For a recent discussion of AD leptogenesis in the context of minimal type II seesaw models, see Ref.~\cite{Barrie:2021mwi}.)  Our goal in this paper is to provide another loop model for neutrino masses, where AD leptogenesis works without adding extra particles and to show how this model also provides a solution to the strong CP problem.

Typically, in the AD mechanism, one relies on the cosmological evolution of a lepton number carrying complex scalar field (called here AD field  and denoted here by $\Phi$), with the Lagrangian of the model explicitly breaking lepton number ($L$), which plays an essential role in the generation of lepton asymmetry. 
While the $L$-breaking term could have any form, we choose it to have a quadratic form in the $\Phi$ field i.e. a $\Phi^2$ term, since with that particular choice, an analytic form for the baryon to entropy ratio can be derived. The neutrino masses in this case arise from the same lepton number breaking $\Phi^2$ term in the Lagrangian. 
 Thus,  neutrino masses are a consequence of AD leptogenesis. 
 Of course, neutrinos in this kind of scenario are naturally Majorana type fermions.   
  There are then restrictions on the parameters of the model following from phenomenological and cosmological consistency. For example, in the AD leptogenesis models, the $L$ asymmetry created by the AD field typically gets transferred to the SM sector at the inflation reheat temperature $T_R$. So any lepton number washout interactions must decouple at temperature $T_*$ with $T_* \gg T_R$. 
Furthermore, one must have $T_R  > T_{sph}$ (where $T_{sph}$ is the sphaleron decoupling temperature) for the lepton asymmetry to be converted to baryon asymmetry. 
While these constraints put a strong restriction on the model parameters, there is still a wide range of them where the model works, as we show below.  


The model in this paper is similar to that of Ref.~\cite{MO1}, though somewhat more economical with the neutrino mass arising from a different diagram.  As in Ref.~\cite{MO1}, we adopt a scheme where the inflaton and the AD fields are one and the same, unlike many original AD scenarios \cite{AD,DRT,mazu1,mazu2}, thus providing unification of inflation and leptogenesis 
\cite{Cline:2019fxx, Charng:2008ke, Hertzberg:2013jba, Takeda:2014eoa, Lin:2020lmr, stubbs, russian, Kawasaki:2020xyf, Barrie:2021mwi, nobu}.  
We find it convenient to adopt the particular scenario proposed in Ref.~\cite{stubbs, russian}, although we believe it can be extended to other types of AD models as well. We include a complex singlet field to implement the PQ solution to the strong CP problem.
One distinguishing feature of our model is that cosmological consistency requires the existence of a near massless Majorana fermion which contributes $\Delta N_{\rm eff}\sim 0.1$ at the Big Bang Nucleosynthesis (BBN) epoch, 
This can be tested in the upcoming CMB-S4 experiment \cite{CMB-S4}. We also note that our model is meant to be a working idea for a unified framework for various puzzles of SM and therefore we do not concern ourselves with the naturalness of the various parameters or their origin from a deeper theory.


This paper is organized as follows: in sec.~2, we present an outline of the model and isolate its symmetries; 
in sec.~3, we discuss the evolution of the universe in this picture, and discuss leptogenesis and  
 one loop generation of neutrino mass in sec.~4;
in sec.~5, we discuss the constraints on the model parameters and provide a benchmark set and in sec.~6, 
dark matter candidate in the model is discussed; 
in sec.~7, we comment on other possible implications of this model. 
Sec.~8 is devoted to a summary of the results.

\section{2. The model} 

The model is based on the SM gauge group $SU(3)_c \times SU(2)_L\times U(1)_Y$.
The particle content is listed in Table I.
In addition to the  SM particle content, we introduce the following new fields i.e.~an AD field $\Phi$, which is an SM singlet scalar and carries a lepton number $-1$, a scalar $SU(2)_L$ doublet $\sigma$  with hypercharge $Y=+1$ and  lepton number $-1$, three Majorana fermionic SM singlets $\chi_i$. To   them, we add the field complex scalar field $ \Delta$, which carries $L=-1$ and the PQ charge $-1$ as in Table I.

\begin{table}[t]
\begin{center}
\begin{tabular}{|c||c||c||c|}
\hline
Field & $U(1)_{PQ}$& SM quantum number & $L$  \\ \hline
Fermion &  &  &\\
$\ell_a$ & $+1$& $({\bf 1},{\bf 2},-1)$ & $+1$ \\
$e^c_a$ & $-1$ & $({\bf 1},{\bf 1}, +2)$ & $-1$\\
$q$&$+1$&$({\bf 3}, {\bf 2}, +1/3$) & $0$\\
$u^c$&$-1$&$({\bf 3^*},{\bf 1},-4/3)$ & $0$\\
$d^c$&$-1$&$({\bf 3^*},{\bf 1},+2/3)$ & $0$\\
$Q$ & $-1$ &$({\bf 3}, {\bf 1},-2/3)$&$+1/2$\\
$Q^c$ & $+2$ & $({\bf 1}, {\bf 3^*},{\bf 1},+2/3)$ & $+1/2$\\
$\chi_i$&$0$ & $({\bf 1}, {\bf 1},0)$& $0$\\\hline
Scalars & &  &\\
$\sigma$ &$-1$ &$({\bf 1}, {\bf 2},+1)$ & $-1$\\
$H$ &$0$ & $({\bf 1},{\bf 2},+1)$ & $0$ \\
$\Phi$ & $+1$& $({\bf 1},{\bf 1},0)$ & $+1$\\
$\Delta$ & $-1$ &$ ({\bf 1},{\bf 1},0)$ & $-1$\\\hline
\hline
\end{tabular}
\caption{
Particle content  of the model responsible for one loop neutrino mass and dark matter and PQ symmetry. 
 $\chi_i$ are new fermionic fields, 
$Q$ and $Q^c$ are new heavy quarks that help implementing the PQ mechanism.
The subscript $a$  goes over lepton flavors and $i$ goes over $\chi$ flavors with $a, i=1,2,3$.
The  PQ charge of the different fields  are shown in the second column. 
 The SM $SU(3)_c \times SU(2)_L\times U(1)_Y$ quantum numbers are in the third column.  
 }
\end{center}
\label{tab:1}
\end{table}

The most general gauge invariant and $U(1)_{PQ}\times U(1)_L$ invariant Lagrangian of the model (in addition to the straightforward kinetic terms) is given symbolically by
\begin{eqnarray}
{\cal L} &=& {\cal L}_{kin} +{\cal L}_{inf} (\Phi, R)-V(\Phi, \Delta,\sigma, H)+ {\cal L}_Y. 
\label{eq:L}
\end{eqnarray}
Here, $ {\cal L}_{Y}$ is the PQ invariant Yukawa Lagrangian given by
\begin{eqnarray}
{\cal L}_{Y}~=~Y_u qHu^c+Y_d q\tilde{H}d^c+Y_\ell \ell \tilde{H}e^c +Y_Q\Delta QQ^c+
(Y_\sigma)_{ai} \ell_a \sigma\chi_i+  \frac{1}{2} \sum_i \mu_{ii} \chi_i\chi_i + h.c.,
 \end{eqnarray}
 and
 \begin{eqnarray}
 V(\Phi,  \Delta, H,\sigma) &=& m^2_\Phi |\Phi|^2+\lambda |\Phi|^4+
 \left(
   \lambda^\prime(\Delta)^2(\Phi^2) +\beta m_\sigma \Phi H^\dagger \sigma  + h.c. \right) \\\nonumber
 && -M^2_{\Delta}|\Delta|^2+\lambda_{\Delta}|\Delta|^4+\lambda_{mix}|\Delta|^2\ |\Phi|^2. 
 \end{eqnarray}
 Here, $\Delta$ is the field whose imaginary part is the axion field. 
${\cal L}_{inf}$ denotes the non-minimal $\Phi$ coupling to gravity of the form 
${\cal L}_{inf}= -\frac{1}{2} (M^2_P + \xi |\Phi|^2) R$ (see, for example, Refs.~\cite{inf1,inf2}) 
and it plays a crucial role in implementing successful inflation, 
 $R$ is the Ricci scalar, and $M_P=2.4 \times 10^{18}$ GeV is the reduced Planck mass.  
 Note that the field $\sigma$ has a lepton number (as does $\Phi$) and $\chi$ being a Majorana fermion has zero lepton number. Without loss of generality, we can work in a basis where the $\chi$ fields are mass eigenstates

We note using Table I that the Lagrangian has an exact global symmetry, $U(1)_{PQ}$ as well as a lepton number symmetry $U(1)_L$. The model also has an automatic $Z_2$ symmetry even after $U(1)_{PQ}$ breaking under which the fields $\Phi, \chi, \sigma$ are odd and the rest of the fields are even.  
This $Z_2$ symmetry remains exact and allows for $\chi_1$ (the lightest among the $Z_2$-odd particles) to be absolutely stable. For subsequent discussion, we assume the following mass hierarchy among the various particles:  
$\mu_{11}, m_H, m_\ell  \ll m_\Phi \leq \mu_{22}, \mu_{33}, m_\sigma$. 
As we will see below, this allows $\Phi$ to decay only via a three body decay mode that involves the field $\chi_1$ in the final state i.e. $\Phi\to \ell_a + \chi_1+H$. As we show below, this will  allow us to relate the reheat temperature $T_R$ directly only to the unknown lightest active neutrino mass, which in turn allows us to choose $T_R$ appropriately.

Once the Field $\Delta$ acquires a vacuum expectation value (vev), it will generate the 
$\epsilon m^2_\Phi\Phi^2$ term, with $\epsilon m^2_\Phi =\lambda^\prime f^2_{PQ}/2$. This term breaks lepton number required for neutrino mass generation as well as for AD leptogenesis. The $\Delta$ vev will also give rise to the axion field which prior to the QCD scale will remain
as a massless particle and solve the strong CP problem. Since $\Phi$ field does not have a vev, its imaginary part  does not contribute to the axion field.

 As we show in a subsequent section, one loop Majorana masses for all neutrinos are proportional to $\epsilon$ whereas the baryon to entropy ratio generated by the AD mechanism is inversely proportional to $\epsilon$ \cite{stubbs,MO1},  thereby relating the neutrino mass with the lepton asymmetry in a way different from traditional leptogenesis. 
 

\section{3.  Inflation and evolution of the AD field } 
To discuss inflation in this model, note that there are two scalar singlets $\Phi$ and $\Delta$ unlike the model in Ref.~\cite{MO1} 
which only had the field $\Phi$ at the epoch of inflation. 
The field $\Delta$ is the mother-field of the axion and implements the PQ symmetry, as already stated above. 
We couple only one of them non-minimally to gravity i.e.
 ${\cal L}_{inf}= -\frac{1}{2} (M^2_P + \xi_\Phi |\Phi|^2) R$. 
To discuss the evolution of the two scalars in the early universe, we expand the fields into the radial and polar parts as 
$\Phi=\frac{1}{\sqrt{2}}\varphi e^{i\theta}$ and $\Delta=\frac{1}{\sqrt{2}}\rho e^{i\delta}$.  
The $\Phi$ part of the potential in the Einstein frame then looks like:
 \begin{eqnarray}
 V_E(\varphi, \rho) \simeq \frac{V(\varphi, \rho)}{ \left(1+\xi\frac{ \varphi^2}{M^2_P} \right)^2}
 \end{eqnarray}
 with
 \begin{eqnarray}
 V(\varphi, \rho ) = \frac{1}{2}m^2_\Phi \varphi^2+\frac{1}{4}\lambda \varphi^4-\frac{1}{2}M^2_\Delta \rho^2+\frac{1}{4}\lambda_\Delta \rho^4
 + \frac{1}{2}\lambda^\prime \rho^2\varphi^2 \cos(2\theta+2\delta) + \frac{1}{4}\lambda_{mix}\rho^2\varphi^2. 
 \end{eqnarray}
 Note the negative sign in front of the $\rho$ mass, which leads to PQ symmetry breaking.
 During inflation, $\varphi \sim M_{P}$ and as a result, the effective potential for $\rho$ turns out to be
 \begin{eqnarray}
 V(\rho) \sim \frac{1}{2}(-M^2_\Delta +\lambda_{mix}M^2_P)\rho^2 +\frac{1}{4}\lambda_\Delta\rho^4 + \frac{\lambda^\prime}{2}M^2_P\rho^2 \cos(2\theta+2\delta). 
 \end{eqnarray}
 We see that by setting $\lambda^\prime \ll \lambda_{mix}$ and $\lambda_{mix} M_P^2 > M_\Delta^2$, 
the mass square of the $\rho$ field is now positive. We therefore expect the $\rho$ field to quickly settle
to its minimum at $ \langle \rho \rangle =0$ and therefore to play no role in inflation or generating curvature fluctuation.

To discuss inflation, we proceed as follows: For $\varphi \geq M_P$, the potential in the Einstein frame is a constant and it leads to inflation. 
As the field $\varphi$ rolls down the potential, its value goes down and 
inflation ends as the slow roll parameters become of order one. 
After that the effect of the coupling of $\varphi$ to the Ricci scalar becomes unimportant. 
The angle $\theta$ can take an arbitrary value when the inflation begins ($\theta ={\cal O}(1)$ is naturally assumed), 
making the real and imaginary parts of the $\Phi$ field different. 
It is this difference which plays a key role in the development of the baryon asymmetry as the $\varphi$ becomes smaller.

After inflation ends, the $\varphi$ field behaves like radiation while the $\varphi^4$ term is dominating the inflaton potential 
and its value goes down like $\varphi\sim a(t)^{-1}$, where $a(t)$ is the scale factor. 
The rest of the story is same as in the paper \cite{stubbs} and concisely explained in Ref.~\cite{nobu}: 
When the $\varphi$ field gets smaller and reaches its value $\varphi_{oscil} \sim {m_\Phi}/\sqrt{\lambda}$, the $\varphi^4$ term becomes unimportant
and the quadratic terms in the Lagrangian dominate $\Phi$ evolution. This leads to a damped harmonic oscillatory behavior 
of the real and imaginary parts of $\Phi$ with different frequencies 
due to the presence of the lepton number breaking term $\epsilon m_\Phi^2 \Phi^2$.
Using the lepton asymmetry formula $n_L\simeq - {\rm Im}(\dot{\Phi} \Phi^*)$, we can then calculate the lepton asymmetry that survives below the reheat temperature. This gives the formula discussed in the next section.  
The only difference between our case and Ref.~\cite{MO1} is the appearance of the $\Delta$ field as an independent field at this temperature. This is because as $\varphi$ becomes smaller, the mass square of the $\Delta$ field becomes negative and PQ symmetry breaks down as $\varphi$ becomes negligible and we get $ \langle \rho \rangle =f_{PQ}=M_\Delta/\sqrt{\lambda_\Delta}$. 
The $\rho$ field then remains stuck there 
and effectively generates the lepton number breaking term $\epsilon m_\Phi^2 \Phi^2$.

To realize the scalar field evolution discussed above, the parameters in the scalar potential must be suitably arranged. 
During inflation, the PQ symmetry is unbroken and hence $M_\Delta^2 < \lambda_{mix} M_P^2$. 
Well before the damped harmonic oscillation of the $\Phi$ field begins, 
$\rho$ must be settled down at $\langle \rho \rangle = f_{PQ}$ to generate the $\epsilon m_\Phi^2 \Phi^2$ term.
This leads to a condition,  
$M_\Delta^2  = \lambda_\Delta f_{PQ}^2 > \lambda_{mix} \varphi_{oscil}^2 \sim (\lambda_{mix}/\lambda) m_\Phi^2$. 
In addition, we impose $\lambda_{mix} f_{PQ}^2 < m_\Phi^2$ in order not to change the formula for the lepton asymmetry
presented in the next section. 
We find that all conditions are easily satisfied. 




\section{4. Lepton asymmetry}

Coming to generation of lepton asymmetry,  we note that the different initial values of the real and imaginary parts of the AD field $\Phi$ i.e. $\phi_1\neq \phi_2$ introduces the CP violation required by the Sakharov's criterion for leptogenesis and leads to lepton asymmetry $n_L= {\rm Im}(\dot{\Phi}^* \Phi) $ while the $\Phi$ field is oscillating. 
This asymmetry gets transmitted to the standard model leptons when $\Phi$ decays to $\ell+\chi_1+H$ and 
reheats the universe to the temperature $T=T_R$.  
There are restrictions on the value of the reheat temperature $T_R$ which imposes constraints on the parameters of the model. 
We must have $T_R \geq T_{sph}\sim 140$ GeV, where $T_{sph}$ is the sphaleron decoupling temperature. 
This is required so that the lepton asymmetry can be converted to the baryon asymmetry. 
Furthermore $T_R < T_*$, where $T_*$ denotes the temperature at which lepton number washout processes 
such as $H \sigma^\dagger \leftrightarrow H^\dagger \sigma$ mediated by $\epsilon$ interaction decouples from the cosmic soup.
  
We estimate the reheat temperature by $T_R\simeq \sqrt{\Gamma_\Phi M_P}$ and find
by using the formula for neutrino mass (see the next section) that it is proportional only to 
the lightest neutrino mass in the normal neutrino mass hierarchy scheme and the latter being unknown at the moment, 
the $T_R$ value can be adjusted as desired.
Here, $\Gamma_\Phi$ is the total decay width of the inflaton/AD field $\Phi$ to  $\ell+\chi_1+H$  (since $m_\sigma > m_\Phi$). 
This part of the discussion is similar to that in Ref.~\cite{MO1}.  We choose $\chi_1$ and $H$ fields  to be lighter than the $\Phi$ field.

We choose parameters such that $T_R =K m_\Phi$ with $K< 1$.  This helps to prevent the inverse decay $\ell+\chi_1+H\to \Phi$ so that the lepton asymmetry generated by $\Phi$ evolution is transmitted to the SM fields.
In the next section, we will see the constraints imposed by this requirement on our model. 

We  first note that in such a leptogenesis scenario, the lepton number to entropy ratio is given by~\cite{stubbs} 
\begin{eqnarray}
\frac{n_L}{s}~\simeq \frac{T_R^3}{\epsilon \, m^2_\Phi \,  M_P}{\rm  sin} 2\theta\simeq 10^{-10}.
\label{nB}
\end{eqnarray}
This formula is valid in our scenario despite the presence of the field $\Delta$ since it gets a vev around $10^{12}$ GeV
and effectively decouples from the $\Phi$ evolution. 

An important input into this estimate of $n_L/s$ is the reheat temperature $T_R=Km_\Phi$, 
which must be less than the AD field mass $m_\Phi$, i.e.~$K< 1$ as already noted. This implies the following relation between $m_\Phi$, $\epsilon $ and $K$ i.e.
\begin{eqnarray}
m_\Phi \simeq 10^{-10}\frac{\epsilon}{K^3} M_P. 
\label{nBsK}
\end{eqnarray}


\section{5.  Neutrino mass, reheat temperature  and washout decoupling}

In this section, we first look at the one loop neutrino mass generation in our model and then its relation to the reheat temperature and the decoupling temperature $T_*$ of the dangerous $L$-violating washout process that could potentially erase the lepton asymmetry. Our main goal will be to establish that in our model, we can satisfy the essential requirement that $T_{sph} \leq T_R \leq T_*$. 
For this purpose, we will assume the following mass hierarchy among the fields, as already stated above,
\begin{eqnarray}
m_\sigma, \mu_{22}, \mu_{33} > m_\Phi \gg \mu_{11}. 
\end{eqnarray}
We will see later on that the $\chi_1$ mass $\mu_{11}$ actually has to be in the eV range or below if it is not to over-close the universe.

 \begin{figure}[tb]
  \centering
 \includegraphics[width=0.7\linewidth]{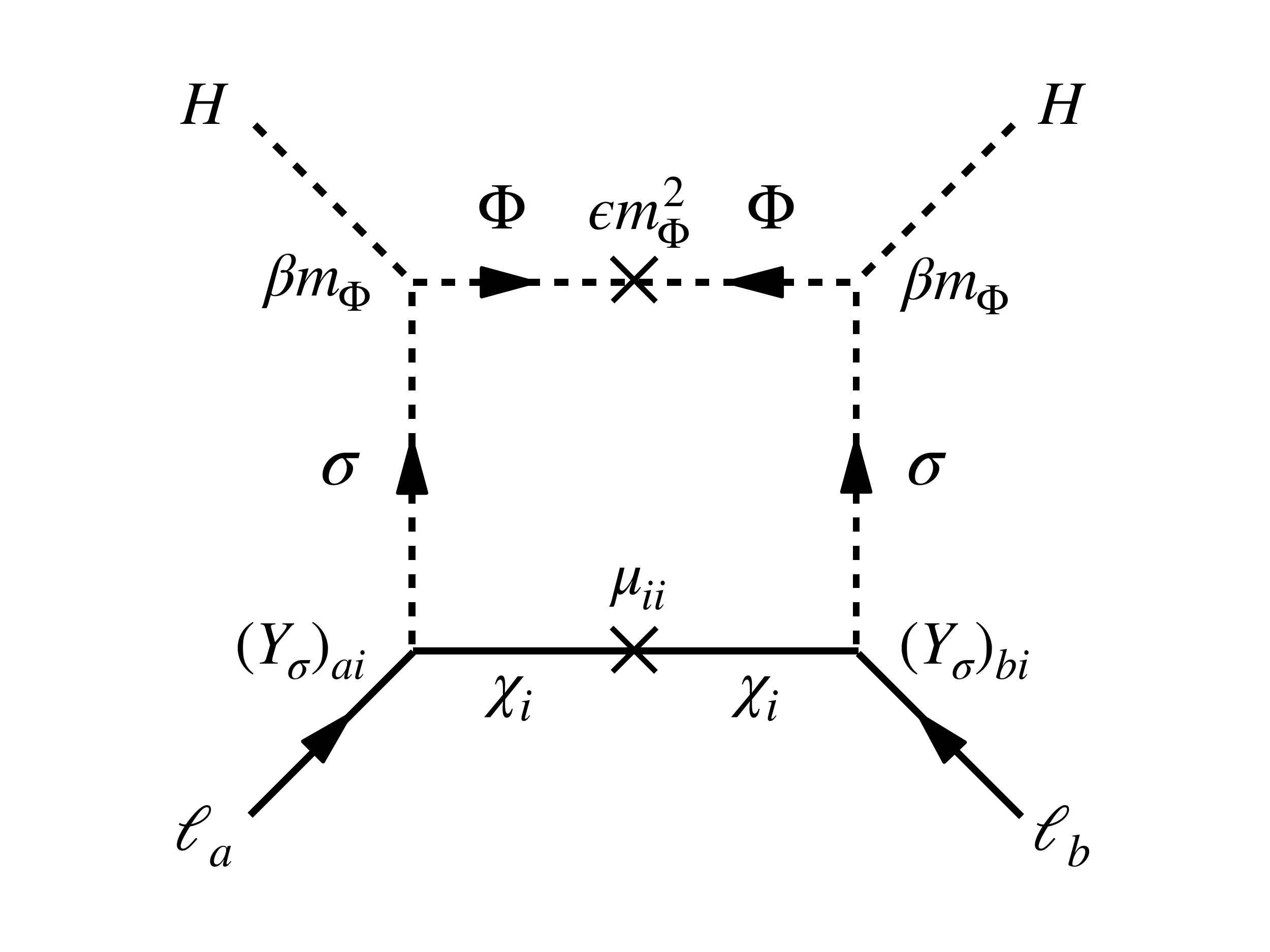}
  \caption{
Feynman diagram responsible for one loop neutrino mass.
Arrows indicate the flow of the lepton number. 
The upper cross denotes the Majorana mass insertion of $(\mu)_{ii}$
while the lower cross is for the insertion of $\epsilon m_\Phi^2$.
} 
  \label{fig1}
  \end{figure}

\noindent{\underline{\bf Neutrino mass}}
  
The diagram for one loop neutrino mass is given in Fig.~1.
We then estimate the light neutrino mass as 
\begin{eqnarray}
m_\nu~=~\frac{ v^2_{wk}\beta^2\epsilon m^2_{\Phi}}{16\pi^2 m^4_\sigma} Y_{\sigma}\mu Y^T_\sigma \equiv X^{-2}Y_\sigma\mu Y_\sigma^T,
\end{eqnarray}
where $X^{-2}=\frac{ v^2_{wk}\beta^2\epsilon m^2_\Phi}{16\pi^2 m^4_\sigma} $, and $\mu={\rm diag}(\mu_{11}, \mu_{22}, \mu_{33})$.
For the second and third generation neutrinos, this one loop result must give a value of ${\cal O}(10^{-10})$ GeV for $m_\nu$. 
It turns out that for $(Y_\sigma)_{2a, 3a}\sim 1$, $\epsilon\sim 10^{-5}$, $\beta \sim 1$, $m_\sigma \sim 10^{6}$ GeV, 
and $m_\sigma = \mu_{22}=\mu_{33} \sim 10^{6.5}$ GeV, we get the correct value for the neutrino masses of second and third generations.
The resulting neutrino masses will then fit the oscillation data. The situation for the lightest neutrino mass $m_{\nu_1}$ is however much smaller as we discuss below. Anyway, the neutrino oscillation fits do not determine the value of $m_{\nu_1}$.

\vskip0.1in

\noindent{\underline{\bf  Reheat temperature and $m_{\nu_1}$}}

Let us now evaluate the reheat temperature in terms of the parameters of the model. For that, we need the decay width of the AD field $\Phi$ 
whose only decay mode is $\Phi\to \ell_a +\chi_1+H$ and it is given by
\begin{eqnarray}
\Gamma_\Phi\simeq \frac{ \beta^2}{32 \pi^3} \, \frac{m_\Phi^5}{m^4_\sigma} \; \sum_a (Y_{\sigma})_{a1}^* (Y_{\sigma})_{a1}.
\label{decay}
\end{eqnarray}
Now using the formula above for neutrino mass, we note that 
$\sum_a  (Y_{\sigma})_{a1}^* (Y_{\sigma})_{a1} 
=X^2\frac{m_{\nu_1}}{\mu_{11}}$, where we have used $m_\nu= U_{MNS}^* D_\nu U_{MNS}^\dagger$
with $D_\nu={\rm diag}(m_{\nu_1}, m_{\nu_2}, m_{\nu_3})$ and the neutrino mixing matrix $U_{MNS}$. 
This leads to the important connection between $T_R$ and $m_{\nu_1}$ i.e. 
\begin{eqnarray}
{T_R}\simeq \frac{\beta x^2}{4\pi\sqrt{2\pi}}X\sqrt{\frac{m_{\nu_{1}}}{\mu_{11}}m_{\Phi}M_P} =\frac{m_\Phi^2}{v_{wk}\sqrt{2\pi \epsilon}} \left(\frac{m_{\nu_1}M_P}{\mu_{11} m_\Phi}\right)^{1/2} ,
\end{eqnarray}
where $x=\frac{m_\Phi}{m_\sigma}$. 
Thus as claimed earlier, this $T_R$ is related to the experimentally undetermined neutrino observable $m_{\nu_1} $ and can be adjusted to satisfy our constraint $T_{sph}\leq T_R < T_*$. Turning this around, we predict $m_{\nu_1}$ for each benchmark choice of parameters to be close to zero. 
For example, when $\epsilon\simeq 10^{-5}$ and $m_\Phi\simeq  10^6$ GeV and $m_\sigma\simeq 10^{6.5}$ GeV, we get $T_R\simeq 10^5$ GeV for $m_{\nu_1}/\mu_{11}\sim 10^{-26}$ while satisfying $n_L/s \sim 10^{-10}$. 
With $\mu_{11}$ in the eV range (as we argue below), $m_{\nu_1}$ is almost massless. 

Note that reheat requires that the mass of one of the three $\chi$ fields must be much lighter than  $\sigma, \Phi $ 
and the Higgs field, as given in Eq.~(\ref{decay}).  
In this case, as we will discuss below, $\chi_1$ decouples from the SM thermal plasma  when it is relativistic and can over-close the universe 
if it is heavier than a few eV (like the neutrino). Therefore, we conclude that $\chi_1$ must have a mass lighter than an eV 
to be cosmologically acceptable. 

There is however no symmetry which guarantees its small mass but nonetheless, we have checked that all loop corrections to its mass are proportional to the neutrino mass and are suppressed, making its small mass technically natural. The leading one loop contribution to $\mu_{11}$ is
\begin{eqnarray}
\delta \mu_{11}\sim \frac{1}{16\pi^2} (m_\nu)_{ab}  (Y_\sigma)_{a1} (Y_\sigma)_{b1}
\frac{\beta^2 \epsilon v^2_{wk}m^2_{\Phi}}{m^4_\sigma}, 
\end{eqnarray}
which is extremely small. 

\vskip0.1in
\noindent{\underline{\bf Washout decoupling temperature $T_*$}} 

Let us now turn our attention to the washout effect from the lepton number violating term $\epsilon m_\Phi^2  \Phi^2$ 
in the effective theory below the PQ symmetry breaking scale.
The rate for  the process $H \sigma^\dagger \leftrightarrow H^\dagger \sigma$ 
which violates lepton number by two units has the potential to wash out any lepton number created. 
If the decoupling temperature of the process ($T_*$) is estimated to be higher than the reheat temperature, 
the washout effect will be absent. Thus we demand that $T_* > T_R$. 

It turns out that when the temperature of the universe is $T \sim m_\sigma$, 
the rate $R$ for the washout process is roughly given by
$R \sim m_\sigma^3 \times \frac{\beta^4 \epsilon^2 m_\Phi^8}{4 \pi m_\sigma^{10}}$. 
If this rate is greater than the Hubble at that time, $H \sim m_\sigma^2/M_P$, 
the washout process is in thermal equilibrium. 
For example, when $\epsilon \simeq 10^{-5}$, $m_\Phi \simeq 10^6$ GeV and $m_\sigma\simeq 10^{6.5}$ GeV, 
we find $R <  H$, so that washout process is out of equilibrium at $T \sim m_\sigma$, 
in other words, $T_* > m_\sigma$ since for temperatures below $m_\sigma$  the Boltzmann suppression 
of $\sigma$ density  keeps the washout process out of equilibrium.  
Since we set $T_R < m_\Phi < m_\sigma$, the lepton asymmetry generated by the AD mechanism does not get washed out.


In Table II, we give two benchmark sets of parameters where the model works. 

\begin{table}[t]
\begin{center}
\begin{tabular}{|c||c||c|}\hline
parameter &value(set 1) &value(set 2)\\\hline
$\epsilon$ & $10^{-5}$ & $10^{-3}$ \\
$K$ & $0.1$ & $0.1$ \\
$m_\Phi$& $10^6$ GeV& $10^8$ GeV \\
$m_\sigma$ & $10^{6.5}$ GeV & $10^{8.5} $ GeV \\
$\beta$ &$\sim 1$  &$\sim 1$ \\
$m_{\chi_1}$ & $\leq 1$ {\rm eV}& $\leq 1$ {\rm eV}  \\
$m_{\nu_1}$ & $\sim 0$ eV & $\sim 0$ eV \\\hline
\end{tabular}
\end{center}
\caption{Two sets of benchmark parameters that satisfy all the constraints considered in the model. 
They cover all points in between and thus represent a broad parameter space of the model.
}
\label{tab2}
\end{table}

\section{6. Prediction of $\Delta N_{\rm eff}$ in the model}
We note from the benchmark parameters given in Table II that the mass of $\chi_1$ fermion is near zero. 
This is required because of the following reason: 
Below $T_R$, the $\chi_1$ is in equilibrium with the SM plasma through $\chi_1 -\ell$ coupling,
and it decouples from the plasma at $T_{D} \sim 1$ TeV (100 TeV) for the choice of benchmark parameters 
$m_{\sigma}\sim 10^{6.5}$ GeV ($10^{8.5}$ GeV).
Thus, the $\chi_1$ field decouples from the thermal plasma when relativistic and as a result the ratio $n_{\chi_1}/n_\gamma$ remains fixed apart from small dilution due to entropy release when other particles annihilate. 
This means that unless the mass of $\chi_1$ is below an eV, it will dominate the energy density (and hence the expansion rate) of the universe, making the theory unacceptable.  
The $\chi_1$ field therefore behaves like a hot dark matter with very small contribution to the universe's energy density $\Omega$. Clearly such a new sub-eV mass particle will leave its imprint on the cosmic microwave background (CMB). 

Using the entropy conservation for the SM plasma and the $\chi_1$ system after the decoupling $T_D$, 
we evaluate the temperature $T_{\chi_1}$ of the $\chi_1$ system at the BBN epoch:   
\begin{eqnarray}
(T_{\chi_1})^3= \frac{g_*^{SM}(T_{BBN})}{g_*^{SM}(T_D)}\, T^3_{BBN}, 
\end{eqnarray}
where $T_{BBN}$ is the temperature of the SM plasma at the BBN ($T_{BBN} \sim 1$ MeV),
and $g_*^{SM}(T)$ is the effective relativistic degrees of freedom of the SM plasma at temperature $T$. 
Since $g^{SM}_*(T\geq 100{\rm GeV})= 106.75$ and  $g^{SM}_*(T_{BBN})= 10.75$, 
we evaluate the extra neutrino species from the $\chi_1$ energy density at the BBN era 
to be $\Delta N_{\rm eff} = 10.75/106.75 \sim 0.1$. 
This is within the reach of the next generation CMB experiment CMB-S4  \cite{CMB-S4} being planned.


\section{7. Comments} We now make several comments on the model:
\begin{itemize}

\item In this model, dark matter is provided by the axion by setting $f_{PQ} \sim 10^{12}$ GeV. 

\item The heavy color triplet field $Q, Q^c$ has mass of order of the PQ breaking scale. 
Although they are super-heavy and stable, they are much heavier than the reheat temperature and therefore are not present 
in the early universe after the reheat when the Hubble phase starts.

\item Due to the presence of only one color triplet fermion coupled to the axion field, the domain wall number $N_{DW}=1$. 
So after the instanton effects kick in there is no domain wall problem.

\item A  prediction of our model is the absence of the right handed neutrinos; so discovery of a right handed neutrino will rule out our model. Similarly, due to the absence of three right-handed neutrinos, our model does not allow for a gauged $B-L$ symmetry~\cite{marshak, david}. So any experimental evidence (see for instance ~\cite{das1, das2, dev})  for a $B-L$ $Z'$ boson would rule out this model.



\item For all our plausible and acceptable scenarios, we find the lightest  active neutrino mass to be close to zero. As a result for this normal mass hierarchy scenario, the neutrinoless double beta decay parameter has a lower limit of  
$\langle m_{\beta\beta} \rangle \geq 0.08$ meV.

\item The model has a near massless Majorana field ($\chi_1$) coupling to leptons. It contributes to $\Delta N_{\rm eff}\simeq 0.1$, 
which can be probed by future precision CMB  experiments such as CMB-S4.
While there is no symmetry which guarantees its tiny mass, we have checked that it is protected from loop corrections being tiny. 

\item The parameter $\lambda^\prime$ that mixes the $\Delta$ and $\Phi$ fields turns out to be very small to give the right order of magnitude for $\epsilon m^2_\Phi$. It becomes bigger as $\Phi$ mass is increased. While we do not address the naturalness issue of parameters in the model here,  we do note that this mixed term is only multiplicatively renormalized due to  quantum corrections and therefore its small value is technically natural. Alternatively, one could envisage  a supersymmetric embedding of the model, where small values of $\lambda^\prime$ will be more natural.

\end{itemize}

\section{8. Summary} 
We have presented an optimal  extension of the standard model that provides a unified explanation of several of its puzzles i.e.~neutrino masses, dark matter compatible with current direct detection constraints, inflation and baryogegenesis via the Affleck-Dine mechanism and a solution to the strong CP problem via the axion. The model adds only three heavy singlet Majorana fermions ($\chi_i$)  to the standard model, supplemented by  a single lepton number carrying a complex SM doublet scalar boson $\sigma$, the singlet lepton number carrying AD field $\Phi$, and a PQ charge carrying field $\Delta$ that implements the strong CP problem solution. All the four features of the model are interconnected: for instance, baryon asymmetry and the neutrino mass are inversely related to each other. The reheat temperature is proportional to the lightest active neutrino mass. 
We give two benchmark points where all the constraints of the model are satisfied.

\section*{Acknowledgement}
The work of R.N.M. is supported by the US National Science Foundation grant no.~PHY-1914631 and  the work of N.O. is supported by the US Department of Energy grant no.~DE-SC0012447.

\end{document}